Effect of anisotropic spin absorption on the Hanle effect in lateral spin valves

H. Idzuchi[1,2*], Y. Fukuma[2,3], S. Takahashi[4,5],

S. Maekawa[5,6] and Y. Otani[1,2*]

[1] *Institute for Solid State Physics, University of Tokyo, Kashiwa 277-8581, Japan*

[2] *Advanced Science Institute, RIKEN, 2-1 Hirosawa, Wako 351-0198, Japan*

[3] *Frontier Research Academy for Young Researchers, Kyushu Institute of Technology, 680-4 Kawazu, Iizuka 820-8502, Japan*

[4] *Institute for Materials Research, Tohoku University, Sendai 980-8577, Japan*

[5] *CREST, Japan Science and Technology, Tokyo 102-0075, Japan*

[6] *Advanced Science Research Center, Japan Atomic Energy Agency, Tokai 319-1195, Japan*

**Abstract**

We have succeeded in fully describing dynamic properties of spin current including the different spin absorption mechanism for longitudinal and transverse spins in lateral spin valves, which enables to elucidate intrinsic spin transport and relaxation mechanism in the nonmagnet. The deduced spin lifetimes are found independent of the contact type. From the transit-time distribution of spin current extracted from the Fourier transform in Hanle measurement data, the velocity of the spin current in Ag with Py/Ag Ohmic contact turns out much faster than that expected from the widely used model.

[*] e-mail address: idzuchi@issp.u-tokyo.ac.jp, yotani@issp.u-tokyo.ac.jp

Rapid development in spintronics is underpinned by solid understanding of fundamental properties of spin transport [1,2]. The dynamic transport properties of spin current have been analyzed by a response of spin precession and dephasing since the pioneering work of Johnson and Silsbee in 1985 [3] and this so-called Hanle effect analysis has been employed up to the present to extract the spin lifetime, the velocity and the transit-time distribution between the injector and the detector [4-9]. However, recent experimental progress in creating spin currents revealed new experimental results which could not be explained by the previous framework. For example, the Hanle analyses of dynamic spin transport properties of graphene, recently performed by assuming an empirical transit-time distribution, yielded strikingly different spin lifetimes depending on the type of contacts although intrinsic (bulk) properties of the spin transport in nonmagnetic materials should be independent of the contact type [8]. In the case of silicon, the experimental Hanle signals could not be fully described by the empirical model based on a drift-dominated transit-time distribution in spin-transport of semiconductor [10,11]. For GaAs, solid analysis of spin relaxation in a two-dimensional electron gas is hampered by complexities of charge and spin transports [12,13]. Therefore, it is essential to provide a framework for understanding the dynamic spin transport properties in the nonmagnetic materials.

In this Rapid Communications, we establish the formalism of Hanle effect to deduce intrinsic spin transport properties in nonmagnetic materials. The experimental studies are based on metallic lateral spin valves, which have comparative advantage in designing the measurement scheme owing to clear physics of charge and spin transport and spin relaxation mechanism [14,15], good controllability of dimensions where one-dimensional transport model is applicable, and comparability of junction property from low resistive transparent junctions to high resistive tunnel junctions [4,7,9,14-21]. As a consequence, we have succeeded in identifying the impact of spin absorption effect on the deduced spin lifetime and obtaining intrinsic spin lifetime which is comparable with other experimental probes such as conduction electron spin resonance.

In order to establish a model of dynamic spin-transport, the Hanle effect was measured in various lateral spin valves (LSVs) with $Ni_{80}Fe_{20}$ (Permalloy, Py)/Ag Ohmic and with Py/MgO/Ag junctions. Samples were prepared on a $Si/SiO_2$ substrate with a suspended resist-mask by using shadow evaporation technique [21] and fabricated LSVs consist of two ferromagnetic Py wires (140-nm-wide and 20-nm-thick [22]) bridged by a nonmagnetic Ag wire (100-140 nm-wide and 100-nm-thick). When the current is applied to the Py/(MgO/)Ag injector junction, the diffusive spin current is generated in the nonmagnetic wire. With the perpendicular magnetic field $B_z$ applied, the spins begin to precess, and the transit time for the spin $t$ is deduced from a change of the angle in the orientation at the detector, which determines the output signal of the device [3,4]. Figure 1 shows Hanle signal for LSVs with both the Py/Ag and Py/MgO/Ag junctions, with the injector-detector separation $L$ varied from 3.00 μm to 6.00 μm. The spin valve signal $\Delta R_S$ corresponds to the difference in non-local resistances between the parallel and antiparallel magnetic configurations of the injector and the detector at $B_Z = 0$. The value of $\Delta R_S$ decreases with increasing $L$ because the spin accumulation decreases due to the spin relaxation in Ag [15]. Also, the values of $\Delta R_S$ for Py/Ag junctions are reasonably smaller than those for Py/MgO/Ag junctions due to the spin resistance mismatch [20,21]: in the case of Ohmic Py/Ag junction, the spin current in the Ag wire is absorbed into Py, which is expected from very low interface resistance $R_I$ for Py/Ag. In Fig. 1, the first cross-point $B_Z^{\pi/2}$ of the Hanle signal for the parallel and antiparallel magnetic configuration of the injector and detector Py wires corresponds to the transit time when the collective $\pi/2$ rotation of diffusive spins is completed. The $B_Z^{\pi/2}$ decreases with increasing $L$ because of the increased transit time in the Ag wire. Figure 1 also shows that the magnitude of $B_Z^{\pi/2}$ alters depending on the type of junctions: for LSV with $L = 6.00$ μm, the Py/Ag junctions give $B_Z^{\pi/2} \sim \pm 156$ mT whereas the Py/MgO/Ag junctions give $\pm 120$ mT. These values correspond respectively to $\omega_L^{\pi/2} \sim 2.75 \times 10^{10}$ s$^{-1}$ and $2.11 \times 10^{10}$ s$^{-1}$, indicating that faster spin diffusion for the Py/Ag junctions compared with the Py/MgO/Ag junctions. This tendency was consistently observed in the LSVs both with $L = 4.50$ μm and $3.00$ μm, the latter of which has the most pronounced difference in $B_Z^{\pi/2}$ between the LSVs with Py/Ag and Py/MgO/Ag junctions.

In order to understand more explicitly the effect of the spin absorption on the dynamic property of spin transport, the transit-time distribution was examined. Hanle signal is described by integrating the transit-time distribution with Larmor precession as

$$V \propto \int_0^\infty dt\, S_y(x=L,t) \equiv \int_0^\infty dt P(t)\cos(\omega_L t), \qquad (1)$$

where $S_y(x=L,t)$ is the net spin density along the $y$ direction parallel to the easy axis of ferromagnet at the detector, $t$ is the transit time and $P(t)$ is the transit-time distribution of the net spin density given by its modulus $S(L,t) = [S_x^2(L,t)+ S_y^2(L,t)]^{1/2}$ [4, 9]. This means that spins injected at $x$=0 arrive at the detector position with a probability of $P(t)$ and the detection voltage is proportional to the integrated $y$-component spin density $S(x=L,t)\cos(\omega_L t)$ with respect to all the possible transit time. After the spin begins to reach the detector, the $P(t)$ increases until the spin-flip nature appears, i.e., the transit time becomes comparable to the spin lifetime. As a result, the transit-time distribution exhibits a typical peak structure as shown in Fig. 2(a), and is usually described by an empirical distribution

$$P_{em}(t) = \frac{1}{\sqrt{4\pi D_N t}} \exp\left[-(L^2/4D_N t)-(t/\tau_{sf})\right], \qquad (2)$$

where $D_N$ is the diffusion constant for spin and $\tau_{sf}$ is the spin lifetime [4]. Considering the fact that the Hanle signal is given by equation (1), $P(t)$ can be directly derived by applying Fourier transform to the experimental Hanle signal [10]. Figure 2(a) and 2(b) show the derived $P(t)$ by performing Fourier transform for the 6 μm spin transport in LSVs. In the case of LSVs with Py/MgO/Ag junctions, experimental data agree excellently with the curve obtained from an empirical model equation (2) with the spin lifetime in table I and the diffusion constant derived with Einstein relation, which validates this scheme. On the other hand, in the case of LSVs with Py/Ag junctions, $P(t)$ from Fourier transform is shifted to the left-hand side with respect to the one expected from the empirical equation (2), suggesting the faster spin diffusion. The experimental $P(t)$ is remarkably different from the empirical equation (2); this makes us desire to construct the model of transit-time distribution to go beyond the empirical one which does not consider the spin absorption.

In order to gain the insight of the effect of spin absorption on the dynamic properties of spin currents in nonmagnet, we formulate the Hanle effect for LSVs with low resistive Ohmic junctions to tunnel junctions. For this, following two issues have to be fully taken into account: firstly, the spin absorption by both injector and detector ferromagnets, affects a spatial distribution of chemical potential [23,24]. In addition to it, a recent experiment of Ghosh *et al* showed that spin relaxation processes in ferromagnets were different between longitudinal and transverse spin currents [25-27]. Their results suggest that the spin relaxation is expected to be more pronounced when the diffusive spins are oriented perpendicular to the magnetization of the detector via precession. The longitudinal component of spin current $I_{Si}^{\parallel}$ through $i$-th junction ($i$ = 1, 2) is described as $I_{Si}^{\parallel} = P_{Ii} G_{Ii} \bar{\mu}_{Fi} / e + (G_{Ii}/2e)[\delta\mu_{Fi} - \delta\mu_{\parallel}(x_i)]$, where $P_{Ii}$ is the interfacial-current spin-polarization, $G_{Ii}$ is the interface conductance, $\bar{\mu}_{Fi} = (\mu_{Fi}^{\uparrow} + \mu_{Fi}^{\downarrow})/2$, $\mu_{Fi}^{\uparrow(\downarrow)}$ is the spin-dependent electrochemical potential of $F_i$, $\delta\mu_{Fi}$ is the spin accumulation of $F_i$ at the interface, $\delta\mu_\parallel(x) = S_y(x)/N(\varepsilon_F)$ is the longitudinal component of spin accumulation in the Ag wire, $N(\varepsilon_F)$ is the density of state at Fermi energy, and $x_i$ is the contact position ($x_1 = 0$, $x_2 = L$). $I_{Si}^\parallel$ is inversely proportional to the spin resistance of $i$-th ferromagnet $R_{Fi}$, as schematically shown in Fig. 3(a). In the presence of transverse spin accumulation $\delta\mu_\perp(x) = S_x(x)/N(\varepsilon_F)$, the transverse spin current $I_{Si}^\perp$ is given by $I_{Si}^\perp = (G_i^{\uparrow\downarrow}/e)\delta\mu_\perp(x_i)$, where $G_i^{\uparrow\downarrow}$ is the real part of spin mixing conductance at the $i$-th interface [28] as schematically shown in Fig. 3(b). The spatial distribution of $\delta\mu_\perp$ and $\delta\mu_\parallel$ are illustrated in Figs. 3(c) and 3(d) with considering different mechanism of spin absorption for longitudinal and transverse spin accumulation, based on the model of Stiles and Zangwill [29]. The spin accumulations, $\delta\mu_\parallel(x)$ and $\delta\mu_\perp(x)$ in the Ag nanowire are given by the complex representation $\delta\tilde{\mu}(x,t) = \delta\mu_\parallel(x) + i\delta\mu_\perp(x)$ [22]

$$\delta\tilde{\mu}(x) = \frac{I_{S1}^\parallel + iI_{S1}^\perp}{2eN(\varepsilon_F)A_N}\int_0^\infty dt P_{em}(x,t)e^{i\omega_L t} + \frac{I_{S2}^\parallel + iI_{S2}^\perp}{2eN(\varepsilon_F)A_N}\int_0^\infty dt P_{em}(x-L,t)e^{i\omega_L t}, \qquad (3)$$

and the spin current density in the complex representation is $\tilde{j}_S(x) = -(\sigma_N/2e)\nabla\delta\tilde{\mu}(x)$, where $\sigma_N$ is the electrical conductivity of Ag wire. Using the boundary conditions that the spin and charge currents are continuous at the interfaces of junctions 1 and 2, we obtain the spin accumulation voltage $V_2 \equiv V$ detected by Py and the nonlocal resistance $V/I$ of Hanle signal in LSV [22], from which parameters can be directly deduced without using effective one [23]. When the junctions are the tunnel junction, the Hanle signal reduces to the conventional expression in LSVs [4,21] in the limit of small spin absorption.

The experimental results are well reproduced by the present theoretical calculations using reasonable parameters listed in Table I, as can be seen in Fig. 1. The obtained spin polarizations $P_F$ and $P_I$ agree well with our previous results [21] and values reported in [30]. The resistivity of Py was $1.75 \times 10^{-5}$ Ωcm. The junction resistance of Py/MgO/Ag was 20 Ω, which is enough higher compared with spin resistance $R_{Ag} = \rho_N \lambda_N / A_N = 1$ Ω. The interfacial resistance of Ohmic Py/Ag junctions and the spin diffusion length of Py are respectively taken as $R_I A_J = 5 \times 10^{-4}$ Ω(μm)² [30] and $\lambda_{Py} = 5$ nm [31] from the literature. $D_N = 612 \pm 19$ cm²/s is derived from Einstein relation $\sigma_N = e^2 D_N N(\varepsilon_F)$ where $N(\varepsilon_F) = 1.55$ states/eV/cm³ [32]. While the shape of Hanle signal is drastically modified by the junctions as in Fig. 1, the spin lifetimes, $40.8 \pm 6.2$ ps and $40.3 \pm 7.3$ ps, for Py/Ag and Py/MgO/Ag junctions agree well with each other. The spin relaxation mechanism is characterized by the spin relaxation ratio $a \equiv \tau_e / \tau_{sf}$ with respect to the momentum relaxation time $\tau_e$. For Ag, $a = 0.10$ ps / 40 ps $= 2.5 \times 10^{-3}$ obtained in this study is consistent with that ($2.50 \times 10^{-3}$) derived from the conduction electron spin resonance (CESR) experiment [33], in which the spin relaxation mechanism was identified as Elliott-Yafet type. The agreement on the value of $a$ for Al and Cu determined from the transport and the CESR measurements has also been reported [3, 14]. Therefore, the spin relaxation time obtained in this study is an intrinsic property of Ag. The characterization was only possible by using the device structure designed in the present study, where the spin transport channel is much longer than the junction size and the surface spin scattering is suppressed by capping layer [15]. In addition to it, the Fourier transform of the theoretical Hanle signal agrees with the experimental $P(t)$ not only for LSVs with Py/MgO/Ag junctions but also for Py/Ag junctions, which complimentary supports the validity of our model. These results show that equation (1) cannot be used with the most widely used $P(t) = P_{em}(t)$ to analyze Hanle signal in LSVs of which $R_I$ is lower than $R_N$ due to the spin absorption effect. They may provide spurious spin lifetimes with mimicking signals or in some cases with different shapes of Hanle signals. In other words, the same spin lifetime results in the different Hanle signal with and without spin absorption, the former of which exhibits a broader signal as shown in Fig. 1. This tendency is consistent with the reported Hanle signals in graphene based LSVs with various type of junctions, where the spin lifetime is deduced as 448-495 ps and 84 ps respectively for tunnel junction and transparent junction [8]. The reanalysis of data using our model provides 448-495 ps and 440 ps for tunnel junctions and transparent junctions, respectively [22], which allows us to separate the intrinsic and extrinsic spin flip mechanisms in graphene.

Spin absorption effect drastically alters the transit-time distribution. The velocity $v$ is estimated as $v \approx L/t_{trans}$ where $t_{trans} \equiv \int_0^\infty dt[tP(t)] / \int_0^\infty dt\, P(t)$. Figures 2(a) and 2(b) show its speed as fast as $9.2 \times 10^4$ m/s for Py/Ag junctions and $6.6 \times 10^4$ m/s for Py/MgO/Ag junctions, which means the diffusion velocity depends on not only diffusion constant but also a spatial gradient in the accumulated spins. The velocity for the Py/Ag junctions is accelerated toward the detector because the spatial distribution of the electrochemical potential is strongly modified by the spin absorption while the diffusion coefficient remains constant in consistent with theoretical report [34]. Figure 2(c) shows $v$ and the full width at half maximum (FWHM) of $P(t)$ for Py/Ag junctions normalized by those for Py/MgO/Ag junctions. For Py/Ag junction not only $v$ is higher but also the FWHM is smaller than those for Py/MgO/Ag junctions, which has the more pronounced difference for short $L$. FWHM is the essential parameter to characterize the coherent spin precession with respect to the applied field because broad distribution of the dwell time gives rise to phase decoherence of the precessing spins [9]. The narrower FWHM for the Ohmic junction may pave the way for efficient control of spins in nonmagnetic material for active spintronic devices.

Our model also enables to derive spin mixing conductance $G_{\uparrow\downarrow}$ which is one of the principal physical quantities characterizing recent novel spintronic effects such as spin pumping and insulating spin Seebeck effect [35,36]. In the present study experimental $G_{\uparrow\downarrow}$ is shown in Table I, whereas theoretical $G_{\uparrow\downarrow}$ is roughly given by Sharvin conductance $G_{\uparrow\downarrow}^{\text{Sh}} = e^2 k_F^2/4\pi h$, where $k_F$ is the Fermi wave number of nonmagnet [37,38]. It provides the value of $G_{\uparrow\downarrow}^{\text{Py/Ag}} \approx G_{\uparrow\downarrow}^{\text{Sh}}$ = $3.7 \times 10^{14}$ $(\Omega m^2)^{-1}$ ($k_F = 1.20 \times 10^{10}$ $m^{-1}$ is from [39]), which is consistent with our experimental values. The larger theoretical value may be due to a reflection of the spin current at the interface [37]. Similar behavior is reported for $G_{\uparrow\downarrow}^{Py/Cu}$ of Py/Cu junctions: the experimental value of $G_{\uparrow\downarrow}^{Py/Cu}$ was obtained as $3.9\times10^{14}$ $(\Omega m^2)^{-1}$ from Giant Magneto Resistance (GMR) study analyzed by circuit theory on ferromagnetic/normal metal hybrid device developed by Brataas *et al* [35,40], which is also smaller than the theoretical value $G_{\uparrow\downarrow}^{Sh} = 4.8 \times 10^{14}$ $(\Omega m^2)^{-1}$ ($k_F = 1.36\times10^{10}$ m$^{-1}$ is from [39]). The quantitative evaluation of $\tau_{sf}$ on the change of $G_{\uparrow\downarrow}$ is shown in [22]. We shall note here that $G_{\uparrow\downarrow}$ obtained in this study is different from the value obtained from spin pumping by a factor of 3-6 [25,41]. For the spin pumping measurement, the magnetization dynamics in ferromagnetic resonance is used for injecting spins in the nonmagnet, and therefore the spin transport properties at the interface may be different from the Hanle effect and GMR measurements using static spin current [42]. The transport parameters generally depend on the frequency i.e. $G=G(\omega)$. Therefore, the Hanle measurements provide us an alternative scheme to determine $G_{\uparrow\downarrow}$.

In summary, we have studied the dynamic transport properties of spin current in metallic lateral spin valves (LSVs) with various junctions. The effect of spin absorption on the Hanle signal was clearly observed in all the devices. The velocity of diffusive spin currents and the transit-time distribution was successfully evaluated by applying Fourier transform to the experimental Hanle signals, resulting in excellent agreement with the empirical model in the case of Py/MgO/Ag junctions. In contrast, we found that the transit-time distribution in LSVs with Py/Ag junctions was strongly deviated from that expected in the empirical model and that the spins diffuse much faster than in LSVs with Py/MgO/Ag junctions, reflecting the spatial distribution of chemical potential affected by the type of junctions. We have successfully formulated the Hanle effect for the LSVs with anisotropic spin absorption for the transverse and longitudinal components of the spin polarization in spin currents relative to the detector magnetization-direction, which enables to elucidate intrinsic spin transport and relaxation mechanisms in the nonmagnet. The model also provides alternative way to determine the spin mixing conductance.

This work was partly supported by Grant-in-Aid for Scientific Research (A) (No. 23244071), (C) (No. 22540346), Young Scientist (A) (No. 23681032) from the Ministry of Education, Culture, Sports, Science and Technology, Japan, and Hoso Bunka Foundation.

**Figure captions**

**Fig. 1.** Hanle signal in LSVs with Py/Ag junctions and Py/MgO/Ag junctions with various separations *L*. Black and red circles show respectively non-local resistance *V*/*I* of parallel and antiparallel magnetic configurations of the injector and detector electrodes at *T* = 10 K. Curves are obtained by the formula of Hanle effect [22] with adjusting parameters shown in Table I. Arrows ($B_z^{\pi/2}$ and $B_z^{-\pi/2}$) show the first cross-points of the Hanle signal for the parallel and antiparallel configurations corresponding to the collective ±*π*/2 rotation of diffusive spins.

**Fig. 2** (a),(b) Derived transit time distribution of pure spin current $P(t)$ (red circle) by performing Fourier transform on Hanle signal shown in Fig. 1(e) and (f). Dashed curves are derived by the empirical model, i.e., diffusion distribution with spin-flip expressed by equation (2), with the values of $D_N$ and $\tau_{sf}$ listed in Table I. Solid curve shows the distribution including the effect of spin absorption [22]. All $P(t)$ is normalized by $P(t_{max})$ where $t_{max}$ gives the maximum of $P(t)$. (c) Velocity and full width at half maximum (FWHM) for spin absorption model normalized by those for empirical model. Lines are guides to the eyes.

**Fig. 3.** (a) Absorbed longitudinal spin current $I_{S\parallel}$ is proportional to longitudinal spin accumulation $\delta\mu_\parallel$ and inversely proportional to the spin resistance of ferromagnet $R_F$. (b) Absorbed transverse spin current $I_{S\perp}$ is proportional to transverse spin accumulation $\delta\mu_\perp$ and the real part of spin mixing conductance $G_{\uparrow\downarrow}$. (c), (d) Schematic of $\delta\mu_\parallel$ and $\delta\mu_\perp$ in the vicinity of the detector junction. In ferromagnet, $\delta\mu_\perp$ is decaying with precessing along the magnetization direction, which results in damping with oscillation [29]. The red and blue curves are calculated by the spin diffusion equation with using the equation (48) in [29].

**Tables**

Table I: Adjusting parameters for Hanle signals which are shown in Fig. 1. The uncertainties of adjusting parameters are determined by the least squares fittings.

| Junction | $L$ (μm) | $P_F$ | $P_{I(Py/MgO/Ag)}$ | $P_{I(Py/Ag)}$ | $\tau_{sf}$ (ps) | $G_{\uparrow\downarrow}$ (m$^{-2}$Ω$^{-1}$) |
|---|---|---|---|---|---|---|
| Py/Ag | 3.00 | 0.57 ± 0.04 | N/A | 0.80 ± 0.03 | 40.3 ± 5.3 | (3.5 ± 0.9) × 10$^{14}$ |
| Py/MgO/Ag | 3.00 | N/A | 0.28 ± 0.02 | N/A | 38.0 ± 3.9 | N/A |
| Py/Ag | 4.50 | 0.51 ± 0.14 | N/A | 0.80 ± 0.10 | 39.3 ± 5.1 | (2.0 ± 0.9) × 10$^{14}$ |
| Py/MgO/Ag | 4.50 | N/A | 0.33 ± 0.05 | N/A | 38.0 ± 6.4 | N/A |
| Py/Ag | 6.00 | 0.55 ± 0.12 | N/A | 0.76 ± 0.06 | 42.9 ± 7.9 | (3.6 ± 8.4) × 10$^{14}$ |
| Py/MgO/Ag | 6.00 | N/A | 0.26 ± 0.07 | N/A | 45.0 ± 10.2 | N/A |

Table I   Idzuchi *et al*.

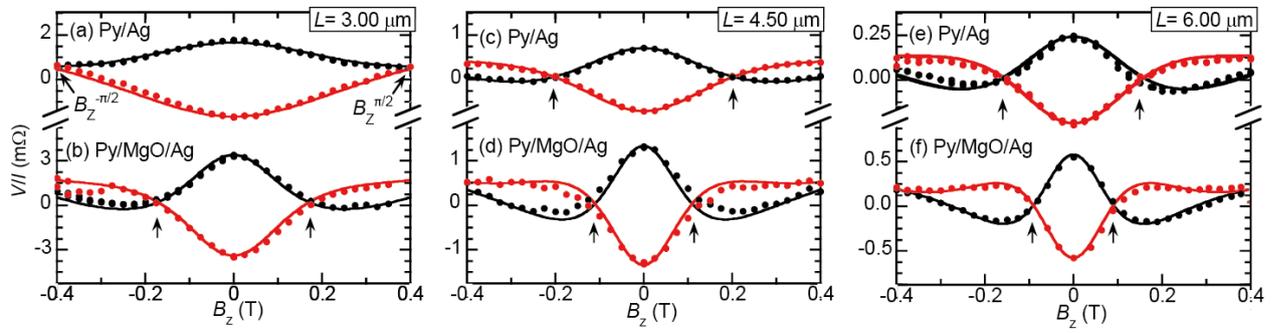

Fig.1   Idzuchi *et al.*

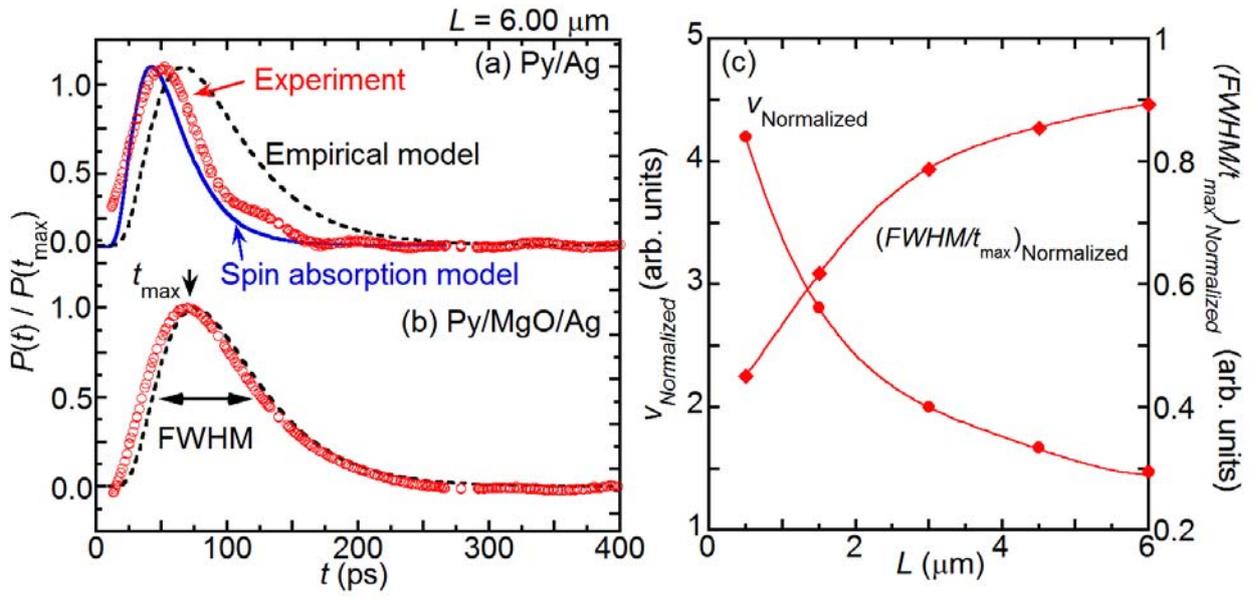

Fig.2　Idzuchi *et al*.

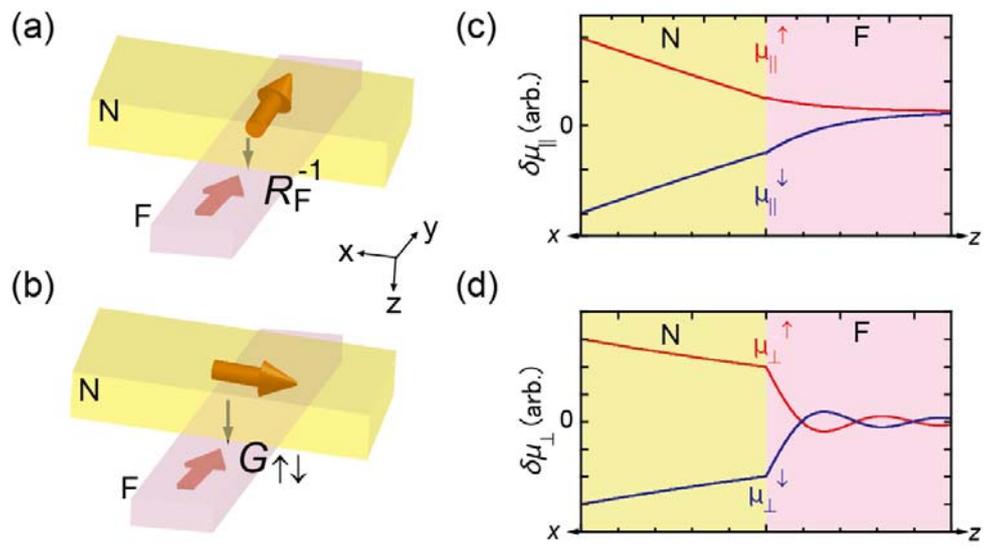

Fig.3  Idzuchi *et al*.

# SUPPLEMENTARY INFORMATION

**Effect of anisotropic spin absorption on the Hanle effect in lateral spin valves**

H. Idzuchi[1,2*], Y. Fukuma[2,3], S. Takahashi[4,5], S. Maekawa[5,6] and Y. Otani[1,2*]

[1]*Institute for Solid State Physics, University of Tokyo, Kashiwa 277-8581, Japan*
[2]*Advanced Science Institute, RIKEN, 2-1 Hirosawa, Wako 351-0198, Japan*
[3]*Frontier Research Academy for Young Researchers, Kyushu Institute of Technology, 680-4 Kawazu, Iizuka 820-8502, Japan*
[4]*Institute for Materials Research, Tohoku University, Sendai 980-8577, Japan*
[5]*CREST, Japan Science and Technology, Tokyo 102-0075, Japan*
[6]*Advanced Science Research Center, Japan Atomic Energy Agency, Tokai 319-1195, Japan*

Nonlocal resistance in a lateral spin valve

When a magnetic field $\mathbf{B} = (0, 0, B_z)$ is applied perpendicular to the plane of a spin injection and detection device consisting of a nonmagnetic metal (N) connected to the ferromagnets of the injector (F1) and the detector (F2) with the magnetizations (white arrows) along the $y$ direction, the injected spins in the N electrode precess around the $z$ axis parallel to $\mathbf{B}$, as shown in Fig. 1. When the spin-current $I_{s1}^i \mathbf{e}_i$ polarized along the $\mathbf{e}_i$ direction ($i = x, y$) is injected from F1 into N at $x = 0$ ($I_{s1}^i > 0$) through the 1st junction and the spin current $I_{s2}^i \mathbf{e}_i$ is absorbed by F2 at $x = L$ ($I_{s2}^i < 0$) through the 2nd junction, the motion of the spin density $\mathbf{S}$ due to the spin accumulation is governed by the diffusion-modified Bloch-Torrey equation [21,43]

$$\frac{\partial \mathbf{S}}{\partial t} = -\gamma_e \mathbf{S} \times \mathbf{B} - \frac{\mathbf{S}}{\tau_{sf}} + D_N \nabla^2 \mathbf{S} + \frac{\hbar}{2e}\frac{I_{s1}^x}{A_N}\mathbf{e}_x \delta(x) + \frac{\hbar}{2e}\frac{I_{s1}^y}{A_N}\mathbf{e}_y \delta(x) \\ + \frac{\hbar}{2e}\frac{I_{s2}^x}{A_N}\mathbf{e}_x \delta(x-L) + \frac{\hbar}{2e}\frac{I_{s2}^y}{A_N}\mathbf{e}_y \delta(x-L), \quad (S1)$$

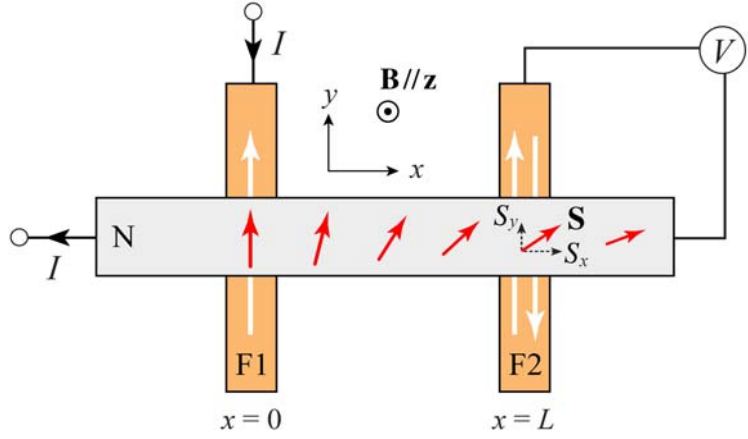

**Fig. S1.** Precession of accumulated spins in N in a lateral spin valve in the presence of perpendicular magnetic field **B** where the spin accumulation (spin density) **S** rotates during the travel of distance $L$ between the injector F1 and the detector F2. The projection of **S** ($S_y$) along the magnetization of F2 is detected by F2 as output voltage $V$.

where $\tau_{sf}$ is the spin-life times, $D_N$ is the spin diffusion constant, and $A_N$ is the cross-sectional area of N electrode, $\mathbf{e}_i$ is the unit vector of the $i$-th direction, and the spin current is taken to be the same unit as charge current. In perpendicular magnetic fields smaller than the demagnetization field, the out-of-plane component $S_z$ of the spin density is small and is disregarded for simplicity. In the steady state ($\partial \mathbf{S}/\partial t = 0$), Equation (S1) is solved to yield the spin density $\mathbf{S} = (S_x, S_y, 0)$ in the complex representation [21]

$$\tilde{S}(x) = S_y(x) + i S_x(x) = \frac{\hbar \tilde{\lambda}_\omega}{4 e D_N A_N} \left[ \tilde{I}_{s1} e^{-|x|/\tilde{\lambda}_\omega} + \tilde{I}_{s2} e^{-|x-L|/\tilde{\lambda}_\omega} \right] \tag{S2}$$

with the complex representation of spin currents $\tilde{I}_{s1} = I^y_{s1} + i I^x_{s1}$ and $\tilde{I}_{s2} = I^y_{s2} + i I^x_{s2}$ through junction 1 and 2 and

$$\tilde{\lambda}_\omega = \frac{\lambda_N}{\sqrt{1+i\omega_L \tau_{sf}}},\qquad (S3)$$

where $\omega_L = \gamma_e B_z$ is the Larmor frequency, $\gamma_e = 2\mu_B/\hbar$ is the gyromagnetic ratio of conduction electrons and $\lambda_N = \sqrt{D_N \tau_{sf}}$ being the spin diffusion length in the absence of the perpendicular magnetic field. Since $\tilde{\lambda}_\omega$ is the complex quantity, $\exp(-|x|/\tilde{\lambda}_\omega)$ exhibits a damped oscillation as a function of $x$. We note that (S2) is rewritten as [44]

$$\tilde{S}(x) = \frac{\hbar}{2e}\frac{\tilde{I}_{s1}}{A_N}\int_0^\infty dt P_{em}(x,t)e^{i\omega_L t} + \frac{\hbar}{2e}\frac{\tilde{I}_{s2}}{A_N}\int_0^\infty dt P_{em}(x-L,t)e^{i\omega_L t},\qquad (S4)$$

where $P_{em}(x,t)$ is the transit-time distribution function:

$$P_{em}(x,t) = \frac{1}{\sqrt{4\pi D_N t}} e^{-x^2/(4D_N t)} e^{-t/\tau_{sf}}.\qquad (S5)$$

When describing the spin transport in the presence of spin precession, it is convenient to use the complex spin accumulation $\delta\tilde{\mu}_N(x) = (2/\hbar)\tilde{S}(x)/N(\varepsilon_F) = \delta\mu_N^y(x) + i\delta\mu_N^x(x)$ given by

$$\delta\tilde{\mu}_N(x) = \frac{e\lambda_\omega}{\sigma_N A_N}\left[\tilde{I}_{s1} e^{-|x|/\lambda_\omega} + \tilde{I}_{s2} e^{-|x-L|/\lambda_\omega}\right],\qquad (S6)$$

where $\sigma_N$ is the electrical conductivity of the N electrode and $N(\varepsilon_F)$ is the density of states (per spin) at the Fermi energy. The absolute value $|\delta\tilde{\mu}_N(x)|$ corresponds to the splitting in the electrochemical potentials (ECP) of the up and down spin electrons. The charge transport is described by the average of the ECPs: $\bar{\mu}_N = -(eI/\sigma_N A_N)x$ for $x < 0$ and $\bar{\mu}_N = 0$ (ground level of ECP) for $x > 0$. In equation (S6), the first term represents the increase of spin accumulation due to spin injection from F1 and the second term is the decrease due to spin absorption by F2. Note that the charge current is absent and the pure spin current flows in the region of $x > 0$. The spin current density flowing in the $x$ direction is given by the complex representation

$$\tilde{j}_N^s(x) = -\frac{\sigma_N}{2e}\nabla\delta\tilde{\mu}_N(x), \tag{S7}$$

Since the thicknesses of the F1 and F2 (~20nm) are much larger than the spin diffusion length $(\lambda_F \sim 5\text{nm})$, we may take the spin-dependent ECPs in F1 and F2 close to the interfaces in the forms of vertical transport along the $z$ direction [24]

$$\mu_{F1}^{\uparrow(\downarrow)}(z) = \mu_{F1}^c - (+)\frac{e\lambda_F}{2\sigma_{F1}^{\uparrow(\downarrow)} A_{I1}}\left(I_{s1}^y - p_{F1}I\right)e^{-z/\lambda_F},$$

$$\mu_{F2}^{\uparrow(\downarrow)}(z) = \mu_{F2}^c - (+)\frac{e\lambda_F}{2\sigma_{F2}^{\uparrow(\downarrow)} A_{I2}} I_{s2}^y e^{-z/\lambda_F}, \tag{S8}$$

where $\mu_{F1}^c = (eI/\sigma_F A_{I1})z + eV_1$ represents the EPC of charge in F1, $A_{Ik}$ is the contact area of the $k$-th interface, $\mu_{F2}^c = eV_2 \equiv eV$ takes a constant potential with no charge current in F2, $V_1$ and $V$ are the voltage drops across junctions 1 and 2, respectively, $\lambda_F$ is the spin-diffusion length of F1 and F2, and $p_{Fk} = (\sigma_{Fk}^\uparrow - \sigma_{Fk}^\downarrow)/\sigma_F$ where $(\sigma_F = \sigma_{Fk}^\uparrow + \sigma_{Fk}^\downarrow)$ is the spin polarization of the F$k$.

The interfacial spin-dependent currents across the $k$-th junction ($k$ = 1, 2) with the polarizations parallel $(\tilde{I}_k^\uparrow)$ and antiparallel $(\tilde{I}_k^\downarrow)$ to the magnetization direction of F1 are [21,24,45,46]

$$\tilde{I}_k^\uparrow = G_{Ik}^\uparrow\left(\frac{\mu_{Fk}^\uparrow(0)}{e} - \text{Re}\left[\frac{\delta\tilde{\mu}_N(x_k)}{2e}\right]\right) - iA_{Ik}G_{Ik}^{\uparrow\downarrow}\,\text{Im}\left[\frac{\delta\tilde{\mu}_N(x_k)}{2e}\right],$$

$$\tilde{I}_k^\downarrow = G_{Ik}^\downarrow\left(\frac{\mu_{Fk}^\downarrow(0)}{e} + \text{Re}\left[\frac{\delta\tilde{\mu}_N(x_k)}{2e}\right]\right) + iA_{Ik}G_{Ik}^{\uparrow\downarrow}\,\text{Im}\left[\frac{\delta\tilde{\mu}_N(x_k)}{2e}\right], \tag{S9}$$

where $G_{Ik}^\sigma$ is the spin-dependent interface conductance of $k$-th junction, $G_{Ik}^{\uparrow\downarrow}$ is the transverse interface conductance per area, so-called the spin-mixing conductance with the dimension $\Omega^{-1}\text{m}^{-2}$, and $x_1 = 0$ and $x_2 = L$. These enable to address the effect of spin absorption not only for longitudinal spin accumulation [21] but also for transverse one. We note that the complex representations (S9) are equivalent to the vector representation in [44]. The total charge and spin currents across the $k$-th interface are $I_k = \tilde{I}_k^\uparrow + \tilde{I}_k^\downarrow$, ($I_1 = I$, $I_2 = 0$) and $\tilde{I}_{sk} = \tilde{I}_k^\uparrow - \tilde{I}_k^\downarrow$. The above interfacial currents are applicable to junctions from tunneling to transparent regime.

Using the boundary conditions that the spin and charge currents are continuous at the interfaces of junctions 1 and 2, we can derive the matrix equation for the interface spin currents

$$\hat{X} \begin{pmatrix} I_{s1}^y \\ I_{s2}^y \\ I_{s1}^x \\ I_{s2}^x \end{pmatrix} = 2 \left( \frac{P_{I1}}{1-P_{I1}^2} \frac{R_1}{R_N} + \frac{P_{F1}}{1-P_{F1}^2} \frac{R_{F1}}{R_N} \right) \begin{pmatrix} I \\ 0 \\ 0 \\ 0 \end{pmatrix}, \tag{S10}$$

with the matrix

$$\hat{X} = \begin{pmatrix} r_{1\parallel} + \mathrm{Re}[\bar{\lambda}_\omega] & \mathrm{Re}[\bar{\lambda}_\omega e^{-L/\tilde{\lambda}_\omega}] & -\mathrm{Im}[\bar{\lambda}_\omega] & -\mathrm{Im}[\bar{\lambda}_\omega e^{-L/\tilde{\lambda}_\omega}] \\ \mathrm{Re}[\bar{\lambda}_\omega e^{-L/\tilde{\lambda}_\omega}] & r_{2\parallel} + \mathrm{Re}[\bar{\lambda}_\omega] & -\mathrm{Im}[\bar{\lambda}_\omega e^{-L/\tilde{\lambda}_\omega}] & -\mathrm{Im}[\bar{\lambda}_\omega] \\ \mathrm{Im}[\bar{\lambda}_\omega] & \mathrm{Im}[\bar{\lambda}_\omega e^{-L/\tilde{\lambda}_\omega}] & r_{1\perp} + \mathrm{Re}[\bar{\lambda}_\omega] & \mathrm{Re}[\bar{\lambda}_\omega e^{-L/\tilde{\lambda}_\omega}] \\ \mathrm{Im}[\bar{\lambda}_\omega e^{-L/\tilde{\lambda}_\omega}] & \mathrm{Im}[\bar{\lambda}_\omega] & \mathrm{Re}[\bar{\lambda}_\omega e^{-L/\tilde{\lambda}_\omega}] & r_{2\perp} + \mathrm{Re}[\bar{\lambda}_\omega] \end{pmatrix}, \tag{S11}$$

where $\bar{\lambda}_\omega = \tilde{\lambda}_\omega / \lambda_N$ and

$$r_{k\parallel} = \left( \frac{2}{1-P_{Ik}^2} \frac{R_{Ik}}{R_N} + \frac{2}{1-p_{Fk}^2} \frac{R_{Fk}}{R_N} \right), \quad r_{k\perp} = \frac{1}{R_N A_{Ik} G_k^{\uparrow\downarrow}}, \quad (k=1,2).$$

Here, $R_{Ik} = 1/G_{Ik}$ ($G_{Ik} = G_{Ik}^\uparrow + G_{Ik}^\downarrow$) is the interface resistance (conductance) of junction $k$, $R_N = (\rho_N \lambda_N / A_N)$ and $R_{Fk} = (\rho_F \lambda_F / A_{Ik})$ are the spin resistances of N and F electrodes, $\rho_N$ and $\rho_F$ are the resistivities, and $P_{Ik} = (G_{Ik}^\uparrow - G_{Ik}^\downarrow)/(G_{Ik}^\uparrow + G_{Ik}^\downarrow)$ is the interfacial spin-current polarization.

The boundary conditions also lead to the nonlocal voltage $V$ due to the spin accumulation detected by F2,

$$V = -\left(\frac{P_{F2}}{1-P_{F2}^2}\frac{R_{F2}}{R_N} + \frac{P_{I2}}{1-P_{I2}^2}\frac{R_{I2}}{R_N}\right) R_N I_{s2}^y, \tag{S12}$$

where the minus sign indicates the absorption of spin current by F2. Using the solution of the matrix equation (S10), we obtain the nonlocal resistance

$$\frac{V}{I} = -2R_N \left(\frac{P_{F1}}{1-P_{F1}^2}\frac{R_{F1}}{R_N} + \frac{P_{I1}}{1-P_{I1}^2}\frac{R_{I1}}{R_N}\right)\left(\frac{P_{F2}}{1-P_{F2}^2}\frac{R_{F2}}{R_N} + \frac{P_{I2}}{1-P_{I2}^2}\frac{R_{I2}}{R_N}\right)\frac{C_{12}}{\det(\hat{X})}, \tag{S13}$$

where $\det(\hat{X})$ is the determinant of the matrix $\hat{X}$ in (S11) and $C_{12}$ is the (1, 2) component of the cofactors of $\hat{X}$,

$$C_{12} = -\det\begin{pmatrix} \operatorname{Re}[\bar{\lambda}_\omega e^{-L/\tilde{\lambda}_\omega}] & -\operatorname{Im}[\bar{\lambda}_\omega e^{-L/\tilde{\lambda}_\omega}] & -\operatorname{Im}[\bar{\lambda}_\omega] \\ \operatorname{Im}[\bar{\lambda}_\omega] & r_{1\perp} + \operatorname{Re}[\bar{\lambda}_\omega] & \operatorname{Re}[\bar{\lambda}_\omega e^{-L/\tilde{\lambda}_\omega}] \\ \operatorname{Im}[\bar{\lambda}_\omega e^{-L/\tilde{\lambda}_\omega}] & \operatorname{Re}[\bar{\lambda}_\omega e^{-L/\tilde{\lambda}_\omega}] & r_{2\perp} + \operatorname{Re}[\bar{\lambda}_\omega] \end{pmatrix}. \tag{S14}$$

When junctions 1 and 2 are tunnel junctions ($R_{Ik} \gg R_N, R_{Fk}$), (S13) reduces to [21,44]

$$\frac{V}{I} = \frac{1}{2} P_{I1} P_{I2} R_N \operatorname{Re}\left[\left(\tilde{\lambda}_\omega/\lambda_N\right) \exp(-L/\tilde{\lambda}_\omega)\right]. \tag{S15}$$

In the absence of perpendicular magnetic field, (S13) reduces to the previous result of [24].

## Simulated Hanle curves for graphene based lateral spin valve

In order to underscore the validity of our analysis, we fit the reported Hanle signal for graphene based lateral spin valve with transparent junctions [8]. As shown in Fig. S2 and table SI, the deduced spin lifetime and diffusion constant are consistent with those from tunnel junctions. $w_F$ = 50 nm, $w_N$ = 2200 nm, $\lambda_F$ = 60 nm, $\rho_F$ = 6 µΩcm, $R_I$ = 285 Ω, $\sigma_N$ = 0.35 mS are taken from [8].

**Table SI:** Adjusting parameters for Hanle signals for graphene based LSV with Co/Graphene junction.

| $L$ (µm) | $P_F$ | $P_{I(Co/Graphene)}$ | $\tau_{sf}$ (ps) | $D_N$ (cm²/s) | $G_{\uparrow\downarrow}$ (m⁻²Ω⁻¹) |
|---|---|---|---|---|---|
| 3.00 | 0.40 | 0.0088 | 440 | 163 | $1.6 \times 10^{10}$ |

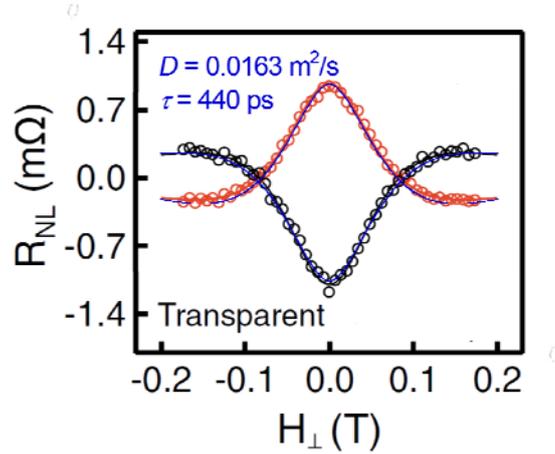

**Fig. S2.** Simulated Hanle curve of graphene based lateral spin valves with transparent junctions. Dots (experimental data) are from [8] and blue lines are calculated from (S13).

**Spin valve measurement**

In order to make the analysis simple, we used the same widths of Py wires. The switching field of each Py wire was controlled by the domain-wall nucleation, i.e., the injector had a large domain wall reservoir at the edge, producing lower switching field than the detector as shown in Fig. S3 [47].

The initial state of the Hanle measurement was set as follows. For the parallel state, firstly Py was initialized by the large field (~ 1000 Oe), and then the field was set to zero. For the antiparallel state, firstly Py was initialized by the large field, and secondly the field was decreased to over the first switching field (~ -100 Oe), finally the field was set to zero.

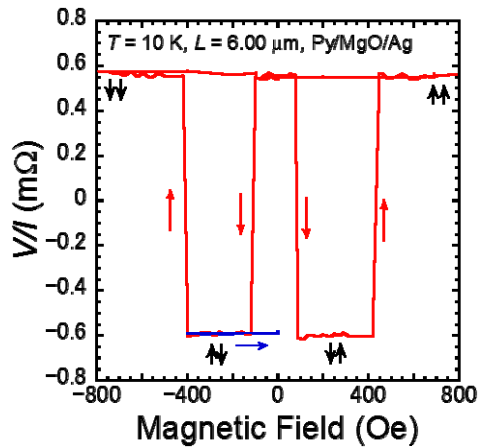

**Fig. S3.** Spin valve measurement (red lines). Blue line indicates the initialization of Py magnetizations for antiparallel configuration. Magnetic field was applied in parallel with the easy axis of Py. Bold arrows show the magnetization states of injector and detector ferro magnet.

**The quantitative evaluation of $\tau_{sf}$ on the change of $G_{\uparrow\downarrow}$**

Here we discuss the effect of anisotropy on the evaluation of spin lifetime. In the case of isotropic spin absorption, $G_{\uparrow\downarrow} = 1/A_J \times \{1/(2R_I)+1/(2R_F)\}$, as shown in (S11). For Ag based LSV with Py/Ag junctions with $L = 3$ μm, the derived $\tau_{sf}$ with isotropic spin absorption is 20 % smaller than the one with anisotropic spin absorption because of underestimation of spin absorption. In contrast, for graphene based LSV with transparent junction ($R_I$ = 285 Ω) [8], the derived $\tau_{sf}$ with isotropic spin absorption is almost same, which is consistent with derived $G_{\uparrow\downarrow}$ is only 4 % different from $1/A_J \times \{1/(2R_I)+1/(2R_F)\}$. The small effect of isotropy is attributed to higher junction resistance compared the one with Ohmic contact in metallic system.

**Additional Reference**

[43] M. Johnson, and R. H. Silsbee, Phys. Rev. B **37**, 5312 (1988).

[44] F. J. Jedema, M. V. Costache, H. B. Heersche, J. J. Baselmans, and B. J. van Wees, Appl. Phys. Lett. **81**, 5162 (2002).

[45] T. Valet, and A. Fert, Phys. Rev. B **48**, 7099 (1993).

[46] X. Wang, G. E. Bauer, B. J. van Wees, A. Brataas, and Y. Tserkovnyak, Phys. Rev. Lett. **97**, 216602 (2006).

[47] H. Idzuchi, Y. Fukuma, L. Wang and Y. Otani, Appl. Phys. Exp. **3**, 063002 (2010).